\documentclass[a4paper,11pt]{article}

\def\be{\begin{equation}}      
  \def\ee{\end{equation}}    \def\beq{\begin{eqnarray}}
  \def\dis{\displaystyle}    \def\eeq{\end{eqnarray}}
       
         \def\ra{\rightarrow}
\usepackage{hyperref}

\begin{document}
\begin{center}
{\Large\bf Decay rates and electromagnetic transitions of heavy quarkonia}

      {\bf J. N. Pandya$^1$, N. R. Soni$^1$, N. Devlani$^2$, A. K. Rai$^3$ }

$^1$ Applied Physics Department, Faculty of Technology \&
Engineering, The M S University of Baroda, Vadodara 390001, Gujarat, INDIA.\\
$^2$ Applied Physics Department, Polytechnic, The M S University of Baroda, Vadodara 390002, Gujarat, INDIA.\\
$^3$ Department of Applied Physics, Sardar Vallabhbhai National Institute of Technology, Surat 395007, Gujarat, INDIA.
\end{center}

\begin{abstract}
The electromagnetic radiative transition widths for heavy quarkonia, as well as digamma and digluon decay widths, are computed in the framework of the extended harmonic confinement model (ERHM) and Coulomb plus power potential (CPP$_\nu$) with varying potential index $\nu$. The outcome is compared with the values obtained from other theoretical models and experimental results. While the mass spectra, digamma and digluon widths from ERHM as well as  CPP$_{\nu=1}$ are in good agreement with experimental data, the electromagnetic transition widths span over a wide range for the potential models considered here making it difficult to prefer a particular model over the others because of the lack of experimental data for most transition widths.
\end{abstract}



\baselineskip=17pt
\section{Introduction}

Decay properties of mesons are of special experimental and theoretical interest because they provide us with further insights into the dynamics of these systems in addition to the knowledge we have gained from the spectra of these families. A large number of experimental facilities worldwide have provided and continue to provide enormous amounts of data which need to be interpreted using the available theoretical approaches \cite{Wiedner2011}. Many phenomenological studies on numerous observables of the $c\bar c$ and $b\bar b$ bound states have established that the non-relativistic nature appears to be an essential ingredient to understand the dynamics of heavy quarkonia \cite{Brambilla:2004}. Thus, the heavy quarkonium spectroscopy is mostly dependent on the quark mass $m$, the momentum $mv$ and  the binding energy $mv^2$ in the non-relativistic limit. Two effective field theories, non-relativistic QCD (NRQCD) \cite{Caswell:1985ui,Bodwin:1994jh}  and potential NRQCD (pNRQCD) \cite{Pineda:1997bj,Brambilla:1999xf}, have been developed leading to a large number of new results for several observables in quarkonium physics \cite{nb2011}.

Radiative transitions in heavy quarkonia have been a subject of interest as the CLEO-c experiment has measured the magnetic dipole (M1) transitions $J/\psi(1S) \rightarrow \gamma \eta_c(1S)$ and $J/\psi(2S) \rightarrow \gamma \eta_c(1S)$ using a combination of inclusive and exclusive techniques and reconciling with theoretical calculations of lattice QCD and effective field theory techniques \cite{rem2009,nb2006}. M1 transition rates are normally weaker than E1 rates, but they are of more interest because they may allow access to spin-singlet states that are very difficult to produce otherwise. It is also interesting that the known M1 rates show serious disagreement between theory and experiment when it comes to potential models. This is in part due to the fact that M1 transitions between different spatial multiplets, such as $J/\psi(1S) \rightarrow \gamma \eta_c(2S \rightarrow 1S)$  are nonzero only due to small relativistic corrections to a vanishing lowest-order M1 matrix element \cite{tb2005}.

We use the spectroscopic parameters of the extended harmonic confinement model (ERHM), which has been successful in predictions of masses of open flavour mesons from light to heavy flavour sectors \cite{pcv1999,jnp2001,jnp2007,akr2008}. The mass spectra of charmonia and bottomonia predicted by this model, and a Coulomb plus power potential (CPP$_\nu$) with varying potential index $\nu$  (from 0.5 to 2.0), employing a non-relativistic treatment for heavy quarks \cite{akr2002,akr2006,akrprc,akr2007}, have been utilized for the present computations along with other theoretical and experimental results.\\

\section{Theoretical framework}

One of the tests for the success of any theoretical model for mesons is the correct prediction of their decay rates. Many phenomenological models predict the masses correctly but overestimate the decay rates \cite{akr2002,akr2006,wb1981}. We have successfully employed a phenomenological harmonic potential scheme and CPP$_\nu$ potential with varying potential index  for different confinement strengths to compute masses of bound states of heavy quarkonia, and the resulting parameters and wave functions have been used to study various decay properties \cite{akr2008}.

The choice of scalar plus vector potential for  quark confinement has been successful in predictions of the low lying hadronic properties in the relativistic  schemes for  quark confinement \cite{sbk1983,skg1987,jbp1996}, which have been extended to accommodate multiquark states from lighter to  heavier flavour sectors with unequal quark masses \cite{pcv1999,jnp2001}. The coloured quarks are assumed to be confined through a Lorentz scalar plus a vector potential of the form
  \be V(r) = \frac{1}{2} (1 + \gamma_0 ) A^2 r^2 + B, \label{eq:potential}\ee
  where $A$ and $B$ are the model parameters and $\gamma_0$ is the Dirac matrix.

  The wave functions for quarkonia are constructed here by  retaining the nature of the single  particle wave function but with a two particle size parameter $\Omega_N({q_iq_j})$,
      \be
  \dis R_{n\ell}(r) =  \dis\left [{\dis{\Omega_N^{3/2}\over{2 \pi}}} {n !\over{\Gamma(n + \ell + {3\over{2}})}}\right ]^{\frac{1}{2}} (\Omega_N^{1/2} r)^\ell \exp\left({-\Omega_N\ r^2\over{2}}\right) L_{n}^{\ell+{1\over{2}}} (\Omega_N r^2) \label{eq:wavefn} \ee
  The Coulombic part of the energy is computed using the residual Coulomb potential using the colour dielectric ``coefficient'', which is found to be state dependent \cite{pcv1999}, so as to get a consistent Coulombic  contribution to the excited states of the
  hadrons. This is a measure of the confinement strength through the non-perturbative
  contributions to the confinement scale at the respective threshold  energies of the quark-antiquark
  excitations.

  The spin average (center of weight) masses of the $c\bar c$ and $b\bar b$ ground states are obtained by choosing the model parameters  $m_c$~=~1.428~GeV,  $m_b$~=~4.637~GeV, $k = 0.1925$ and  the confinement parameter $A = 0.0685$ GeV$^{3/2}$ \cite{pcv1999,jnp2001}.

  In the other approach using the CPP$_\nu$ scheme for the heavy-heavy bound state systems such as $c \bar c$ and $b \bar b$, we treat the motion of both the quarks and antiquarks nonrelativistically \cite{akr2008}. The CPP$_\nu$ potential is given by
\be V(r)=\frac{-\alpha_c}{r} + A r^\nu
\label{eq:405}\ee
Here, for the study of heavy flavoured mesons, $\alpha_c=4\alpha_s/3$, $\alpha_s$
being the strong running coupling constant, $A$ is the potential parameter and
$\nu$ is a general power, such that the choice $\nu= 1$ corresponds to the Coulomb plus linear potential.

We have employed the hydrogenic trial wave function here for the present calculations. For excited states we consider the wave function multiplied by an appropriate orthogonal polynomial function such that the generalized variational wave function gets orthonormalized. Thus, the  trial wave function for the $(n,l)$ state is assumed to be the form given by
\begin{equation}  \label{eq:wavfun}
R_{nl}(r) = \left(\frac{\mu^3 (n-l-1)!}{2n(n+l)!}\right)^{\frac{1}{2}}
(\mu \ r)^l \ e^{- \mu r /2} L^{2l+1}_{n-l-1}(\mu r).\end{equation}
Here, $\mu$ is the variational parameter and $L^{2l+1}_{n-l-1}(\mu
r)$ is a Laguerre polynomial.\\
For a chosen value of $\nu$, the variational parameter $\mu$ is
determined for each state using the virial theorem
\begin{equation}
\label{eq:virial}
 \left<KE\right>=\frac{1}{2}\left<\frac{r
 d V}{dr}\right>.\end{equation}
The potential index $\nu$ is chosen to vary from $0.5$ to $2$. Quark mass parameters are fitted to get the experimental ground state masses of $ m_c=1.31 \ GeV$, $m_b = 4.66 \ GeV$, $\alpha_c=0.4$ (for $c\bar c $)  and $\alpha_c=0.3$ (for $b\bar b $). The potential parameter $A$ also varies with $\nu$ \cite{akrprc}.

We have done a completely parameter-free computation of digamma and digluon decay widths and radiative electric and magnetic dipole transition widths using the parameters of these phenomenological models that were fixed to obtain the ground state masses of the quarkonia systems.

\begin{table*}[t]
\center
\caption{Digamma decay width of charmonia (keV)}\label{tab:ccdigamma}
\begin{tabular}{lccccccccccc}
\hline\hline
                      &  $1^1S_0$ & $2^1S_0$ & $3^1S_0$ & $4^1S_0$ & $1^3P_0$ &  $1^3P_2$ &  $2^3P_0$ &  $2^3P_2$ \\
\hline
$ERHM$                & 8.76      & 5.94     & 3.05     & 1.43     & 69.97    & 73.93     & 6.93      & 6.98   \\
$ERHM(corr)$          & 6.21      & 4.21     & 2.17     & 1.01     & 71.04    & 75.06     & 5.87      & 5.91   \\
CPP$_{\nu=0.5}$       & 12.85     & 3.47     & 1.83     & 1.24     & 5.74     & 1.54      & 21.11     & 5.69   \\
CPP$_{\nu=0.5}(corr)$ & 7.32      & 1.98     & 1.04     & 0.71     & 5.84     & 1.19      & 21.59     & 4.40   \\
CPP$_{\nu=1.0}$       & 22.79     & 9.88     & 6.73     & 5.28     & 27.29    & 7.45      & 143.30    & 39.41  \\
CPP$_{\nu=1.0}(corr)$ & 12.99     & 5.63     & 3.84     & 3.01     & 27.91    & 5.76      & 146.57    & 30.49  \\
CPP$_{\nu=1.5}$       & 30.84     & 17.55    & 14.16    & 12.65    & 63.35    & 17.52     & 511.88    & 144.33 \\
CPP$_{\nu=1.5}(corr)$ & 17.58     & 10.00    & 8.07     & 7.21     & 64.79    & 13.56     & 523.53    & 111.66 \\
CPP$_{\nu=2.0}$       & 37.43     & 25.11    & 22.88    & 22.43    & 108.06   & 30.26     & 1058.7    & 305.98 \\
CPP$_{\nu=2.0}(corr)$ & 21.34     & 14.31    & 13.04    & 12.79    & 110.52   & 23.41     & 1082.8    & 236.72 \\
\cite {manan2012}     & 10.38     & 3.378    & 1.9      & 1.288    & --       & --        & --        & --\\
\cite{bql2009}        & 8.5       & 2.4      & 0.88     &  --      & 2.5      & 0.31      & 1.7       & 0.23   \\
\cite{schuler1998}    & 7.8       & 3.5      & --       & --       & --       & --        & --        & --\\
\cite{ahmady1995}     & 11.8      & --       & --       & --       & --       & --        & --        & -- \\
\hline\hline
\end{tabular}
\end{table*}

\begin{table*}
\center
\caption{Digluon decay width of charmonia (MeV)}\label{tab:ccdigluon}
\begin{tabular}{lccccccccccc}
\hline\hline
                       &  $1^1S_0$    & $2^1S_0$  & $3^1S_0$ & $4^1S_0$ & $1^3P_0$     & $1^3P_2$       & $2^3P_0$  & $2^3P_2$ \\
\hline
$ERHM$                 & 13.48        & 9.14      & 4.7      & 2.19     & 0.11         & 0.11           & 9.07     & 9.13 \\
$ERHM(corr)$           & 19.04        & 12.91     & 6.64     & 3.1      & 0.19         & 0.2            & 5.31      & 5.43 \\
CPP$_{\nu=0.5}$        & 43.41        & 11.73     & 6.17     & 4.19     & 0.019        & 3.71           & 0.07      & 13.74\\
CPP$_{\nu=0.5}(corr)$  & 69.94        & 18.89     & 9.94     & 6.76     & 0.040        & 1.43           & 0.15      & 5.29\\
CPP$_{\nu=1.0}$        & 77.01        & 33.37     & 22.74    & 17.84    & 0.092        & 17.99          & 0.48      & 95.21\\
CPP$_{\nu=1.0}(corr)$  & 124.08       & 53.77    & 36.64    & 28.74    & 0.195        & 6.93           & 1.02     & 36.69\\
CPP$_{\nu=1.5}$        & 104.18       & 59.28     & 47.85    & 42.73    & 0.214        & 42.33          & 1.73       & 348.66\\
CPP$_{\nu=1.5}(corr)$  & 167.85       & 95.51     & 77.09    & 68.85    & 0.453        & 16.31          & 3.66     & 134.38\\
CPP$_{\nu=2}$          & 126.46       & 84.83     & 77.29    & 75.79    & 0.365        & 73.11          & 3.58     & 739.15\\
CPP$_{\nu=2.0}(corr)$  & 203.75       & 136.67    & 124.53   & 122.12    & 0.773         & 28.18         & 7.57     & 284.88\\
\cite{pdg2014}         & 26.7$\pm$3.0 & --        & --       & --       & 10.2$\pm$0.7 & 2.034$\pm$0.12 & --        & --\\
\cite{ap2010}          & 48.927       & --        & --       & --       & 38.574       & 4.396          & --        & -- \\
\cite{laverty}pert.    & 15.70        & --        & --       & --       & 4.68         & 1.72           & --        & --\\
\cite{laverty}nonpert. &10.57         & --        & --       & --       & 4.88         & 0.69           & --        & --\\
\hline\hline
\end{tabular}
\end{table*}

\begin{table*}
\center
\caption{Digamma decay width of bottomonia (keV)}\label{tab:bbdigamma}
\begin{tabular}{lccccccccccc}
\hline\hline
  &  $1^1S_0$ &  $2^1S_0$ &  $3^1S_0$ &  $4^1S_0$ &  $1^3P_0$  &  $1^3P_2$ &  $2^3P_0$ &  $2^3P_2$ \\
\hline
$ERHM$                 & 0.47  & 0.26  & 0.12  & 0.01    & 1.37   & 1.39   & 0.12   & 0.12 \\
$ERHM(corr)$           & 0.35  & 0.20  & 0.09  & 0.07    & 1.39   & 1.40   & 0.10   & 0.10 \\
CPP$_{\nu=0.5}$        & 0.36  & 0.06  & 0.03  & 0.038   & 0.02   & 0.005  & 0.057  & 0.015\\
CPP$_{\nu=0.5}(corr)$  & 0.24  & 0.04  & 0.02  & 0.026   & 0.02   & 0.004  & 0.058  & 0.013\\
CPP$_{\nu=1.0}$        & 0.55  & 0.15  & 0.09  & 0.080   & 0.08   & 0.022  & 0.42   & 0.11\\
CPP$_{\nu=1.0}(corr)$  & 0.37  & 0.10  & 0.06  & 0.054   & 0.08   & 0.018  & 0.43   & 0.09\\
CPP$_{\nu=1.5}$        & 0.71  & 0.27  & 0.18  & 0.123   & 0.20   & 0.055  & 1.34   & 0.36\\
CPP$_{\nu=1.5}(corr)$  & 0.48  & 0.18  & 0.12  & 0.084   & 0.21   & 0.045  & 1.36   & 0.30\\
CPP$_{\nu=2.0}$        & 0.84  & 0.38  & 0.29  & 0.165   & 0.35   & 0.095  & 2.83   & 0.76\\
CPP$_{\nu=2.0}(corr)$  & 0.57  & 0.26  & 0.20  & 0.112   & 0.36   & 0.078  & 2.88   & 0.63\\
\cite {manan2012}      & 0.496 & 0.212 & 0.135 & 0.099   & --     & --     & --     & --\\
\cite{bql2009}         & 0.527 & 0.263 & 0.172 & --      & 0.037  & 0.0066 & 0.037  & 0.0067\\
\cite{schuler1998}     & 0.460 & 0.20  & --    & --      & --     & --     & --     & --\\
\cite{ahmady1995}      & 0.580 & --    & --    & --      & --     & --     & --     & --\\
\hline\hline
\end{tabular}
\end{table*}

\begin{table*}
\center
\caption{Digluon decay width of bottomonia (MeV)}\label{tab:bbdigluon}
\begin{tabular}{lccccccccccc}
\hline\hline
  &  $1^1S_0$ &  $2^1S_0$ &  $3^1S_0$ &  $4^1S_0$ &  $1^3P_0$  &  $1^3P_2$ &  $2^3P_0$ &  $2^3P_2$ \\
\hline
$ERHM$                      & 7.61  & 4.31   & 1.99  & 1.58   & 22.45  & 22.68  & 1.93  & 1.94 \\
$ERHM(corr)$                & 9.95  & 5.64   & 2.61  & 2.07   & 38.17  & 38.57  & 1.92  & 1.92 \\
CPP$_{\nu=0.5}$             & 10.92 & 1.77   & 0.78  & 1.17   & 0.61   & 0.16   & 1.74  & 0.46\\
CPP$_{\nu=0.5}(corr)$       & 15.51 & 2.51   & 1.11  & 1.66   & 1.20   & 0.16   & 3.40  & 0.46\\
CPP$_{\nu=1.0}$             & 16.71 & 4.65   & 2.72  & 2.43   & 2.51   & 0.67   & 12.81 & 3.42\\
CPP$_{\nu=1.0}(corr)$       & 23.72 & 6.61   & 3.86  & 3.45   & 4.90   & 0.66   & 25.04 & 3.39\\
CPP$_{\nu=1.5}$             & 21.53 & 8.14   & 5.60  & 3.76   & 6.22   & 1.67   & 40.70 & 10.91\\
CPP$_{\nu=1.5}(corr)$       & 30.58 & 11.55  & 7.95  & 5.34   & 12.16  & 1.65   & 79.57 & 10.81\\
CPP$_{\nu=2.0}$             & 25.55 & 11.66  & 8.95  & 5.03   & 10.74  & 2.88   & 86.12 & 23.15\\
CPP$_{\nu=2.0}(corr)$       & 36.29 & 16.56  & 12.72 & 7.14   & 21.00  & 2.85   & 168.36& 22.93\\
\cite{ap2010}               & 14.64 & --     & --    & --     & 2.745  & 0.429  & --    & --\\
\cite{laverty}pert.         & 11.49 & --     & --    & --     & 0.96   & 0.33   & --    &--\\
\cite{laverty}nonpert.      & 12.39 & --     & --    & --     & 2.74   & 0.25   & --    &--     \\
\cite{gupta1996}            & 12.46 & --     & --    & --     & 2.15   & 0.22   & --    &--\\
\hline\hline
\end{tabular}
\end{table*}

\begin{table*}
\caption{E1 transition partial widths of $c\bar c$ (keV)}\label{tab:cce1}
\begin{center}
\begin{tabular}{cccccccccccc}
  \hline\hline
  Transitions&ERHM&\multicolumn{4}{c}{CPP$_{\nu}$}&\cite{de2003}&\cite{sfr2007}& \cite{ap2010} & \cite{nby2005}&\cite{bql2009}&\cite{pdg2014}\\
  \hline
  && $_{\nu=0.5}$ & $_{\nu=1.0}$ & $_{\nu=1.5}$ & $_{\nu=2.0}$&&& &&&\\
  \hline
  $ 2^3S_1 \ra1^3P_0 $ & 9.2   & 6.7  & 38.2  & 89.2  & 145.8   & 51.7 & 45    & --    & 47    & 74    & 29.8$\pm$1.29\\
  $ 2^3S_1 \ra1^3P_1 $ & 18.6  & 13.8 & 73.6  & 164.6 & 259.7   & 44.9 & 40.9  & --    & 42.8  & 62    & 28.2$\pm$1.47\\
  $ 2^3S_1 \ra1^3P_2 $ & 11.3  & 8.4  & 37.2  & 72.4  & 100.3   & 30.9 & 26.5  & --    & 30.1  & 43    & 26.5$\pm$1.3\\
  $ 3^3S_1 \ra2^3P_0 $ & 16.4  & 5.9  & 51.4  & 164.3 & 349.2   & --   & 87.3  & --    & --    & --    & \\
  $ 3^3S_1 \ra2^3P_1 $ & 43.3  & 8.4  & 65.2  & 192.7 & 382.9   & 65.7 & --    & --    & --    & --    & \\
  $ 3^3S_1 \ra2^3P_2 $ & 54.2  & 1.6  & 4     & 4.1   &3.1      & --   & 31.6  & --    & --    & --    & \\
  $ 3^3S_1 \ra1^3P_0 $ & 129.4 & 105.1& 583.9 & 1389  & 2274    & --   & 1.2   & --    & --    & --    & \\
  $ 3^3S_1 \ra1^3P_1 $ & 336.4 & 281.5& 1531  & 3607  & 5863    & --   & 2.5   & --    & --    & --    & \\
  $ 3^3S_1 \ra1^3P_2 $ & 410.1 & 1897 & 4379  & 6998  & --      & --   & 3.3   & --    & --    & --    & \\
  $ 1^3P_2\ra 1^3S_1 $ & 680.7 & 168  & 421   & 652   & 828     & 448  & 390.6 & 250   & 315   & 424   & 390$\pm$26\\
  $ 1^3P_1\ra 1^3S_1 $ & 426.2 & 127  & 269   & 363   & 409     & 333  & 287   & 229   & 41    & 314   & 299$\pm$22\\
  $ 1^3P_0\ra 1^3S_1 $ & 325.9 & 110  & 209   & 256   & 264     & 161  & 142   & 173   & 120   & 152   & 133$\pm$9\\
  $ 1^1P_1\ra 1^1S_0 $ & 1076.2& 401  & 1015  & 1569  & 2000    & 723  & 610   & 451   & 482   & 498   & \\
  $ 2^3P_2\ra 2^3S_1 $ & 325.3 & 151  & 701   & 1707  & 2883    & --   & 358.6 & 83    & --    & 225   & \\
  $ 2^3P_1\ra 2^3S_1 $ & 258.9 & 92   & 316   & 596   & 824     & --   & 208.3 & 73.8  & --    & 103   & \\
  $ 2^3P_0\ra 2^3S_1 $ & 231.0 & 68   & 190   & 291   & 322     & --   & 53.6  & 49.4  & --    & 61    & \\
  $ 2^1P_1\ra 2^1S_0 $ & 611.7 & 184  & 843   & 1961  & 3219    & --   & --    & 146.9 & --    & 309   & \\
  $ 2^3P_2\ra 1^3S_1 $ & 700.1 & 187  & 1279  & 3510  & 5896    & --   & 33    & 140   & --    & 101   & \\
  $ 2^3P_1\ra 1^3S_1 $ & 661.3 & 160  & 962   & 2352  & 3590    & --   & 28    & 133   & --    & 83    & \\
  $ 2^3P_0\ra 1^3S_1 $ & 643.5 & 146  & 822   & 1880  & 2683    & --   & 21    & 114   & --    & 74    & \\
  $ 2^3P_1\ra 1^1S_0 $ & 951.6 &  93  & 549   & 1321  & 2013    & --   & --    & 227   & --    & 134   & \\
  \hline\hline
  \end{tabular}
  \end{center}
  \end{table*}

\begin{table*}
\caption{E1 transition partial widths of $b\bar b$ (keV)}\label{tab:bbe1}
\begin{center}
\begin{tabular}{cccccccccccc}
  \hline\hline
 Transitions &ERHM&\multicolumn{4}{c}{CPP$_{\nu}$}&\cite{de2003}&\cite{sfr2007}& \cite{ap2010} & \cite{nby2005}&\cite{bql2009}&\cite{pdg2014}\\
  \hline
  && $_{\nu=0.5}$ & $_{\nu=1.0}$ & $_{\nu=1.5}$ & $_{\nu=2.0}$&&& &&&\\
  \hline
  $ 2^3S_1 \ra1^3P_0 $ & 0.24    & 0.06  & 0.4       & 1.08  & 1.63  & 1.65  & 1.15  & --    & 1.29  & 1.67  & 1.21$\pm$0.16\\
  $ 2^3S_1 \ra1^3P_1 $ & 0.40    & 0.12  & 0.74      & 1.75  & 2.71  & 2.57  & 1.87  & --    & 2.0   & 2054  & 2.21$\pm$0.22\\
  $ 2^3S_1 \ra1^3P_2 $ & 0.12    & 0.04  & 0.38      & 1.39  & 3.03  & 2.53  & 1.88  & --    & 2.04  & 2.62  & 2.29$\pm$0.22\\
  $ 3^3S_1 \ra2^3P_0 $ & 0.35    & 0.04  & 0.32      & 1.03  & 2.16  & 1.65  & 1.67  & --    & 1.35  & 1.83  & 1.2$\pm$0.16\\
  $ 3^3S_1 \ra2^3P_1 $ & 0.82    & 0.08  & 0.62      & 1.78  & 3.60  & 2.65  & 2.74  & --    & 2.20  & 2.96  & 2.56$\pm$0.34\\
  $ 3^3S_1 \ra2^3P_2 $ & 0.80    & 0.06  & 0.30      & 0.62  & 0.98  & 2.89  & 2.80  & --    & 2.40  & 3.23  & 2.66$\pm$0.41\\
  $ 3^3S_1 \ra1^3P_0 $ & 3.91    & 2.38  & 15.4      & 40.4  & 72.0  & 0.124 & 0.03  & --    & 0.001 & 0.07  & $0.055\pm0.08$\\
  $ 3^3S_1 \ra1^3P_1 $ & 9.50    & 6.38  & 41.1      & 106.8 & 188.8 & 0.307 & 0.09  & --    & 0.008 & 0.17  & $<0.018\pm0.001$\\
  $ 3^3S_1 \ra1^3P_2 $ & 9.86    & 8.22  & 54.7      & 153.7 & 290.8 & 0.445 & 0.13  & --    & 0.015 & 0.25  & $<0.2\pm0.32$\\
  $ 1^3P_2\ra 1^3S_1 $ & 61.96   & 11.3  & 26.7      & 40.1  & 48.8  & 42.7  & 31    & 44.0  & 31.6  & 38    & \\
  $ 1^3P_1\ra 1^3S_1 $ & 39.58   & 09.4  & 21.3      & 33.3  & 43.5  & 37.1  & 27    & 42.0  & 27.8  & 34    & \\
  $ 1^3P_0\ra 1^3S_1 $ & 30.72   & 08.6  & 18.7      & 27.8  & 35.0  & 29.5  & 22    & 37.0  & 22.0  & 27    & \\
  $ 1^1P_1\ra 1^3S_0 $ & 62.70   & 15.7  & 37.7      & 60.4  & 81.6  & --    & 38    & 60.0  & --    & 56.8  & \\
  $ 2^3P_2\ra 2^3S_1 $ & 14.57   & 04.9  & 23.4      & 55.5  & 96.1  & 18.8  & 17    & 20.4  & 14.5  & 18.8  & \\
  $ 2^3P_1\ra 2^3S_1 $ & 10.65   & 04.3  & 18.2      & 39.5  & 63.7  & 15.9  & 14    & 12.5  & 12.4  & 15.9  & \\
  $ 2^3P_0\ra 2^3S_1 $ & 8.98    & 03.9  & 15.9      & 32.8  & 51.1  & 11.7  & 10    & 4.4   & 9.2   & 11.7  & \\
  $ 2^1P_1\ra 2^1S_0 $ & 15.67   & 05.4  & 25.4      & 60.0  & 102.1 & 23.6  & --    & 25.8  & --    & 24.7  & \\
  $ 2^3P_2\ra 1^3S_1 $ & 45.03   & 09.0  & 33.0      & 67.2  & 104.0 & 8.41  & 7.74  & 20.8  & 12.7  & 13    & \\
  $ 2^3P_1\ra 1^3S_1 $ & 41.71   & 08.6  & 30.2      & 58.9  & 88.0  & 8.01  & 7.31  & 19.9  & 12.7  & 12.4  & \\
  $ 2^3P_0\ra 1^3S_1 $ & 40.12   & 08.4  & 28.8      & 55.0  & 80.8  & 7.36  & 6.69  & 14.1  & 10.9  & 11.4  & \\
  $ 2^1P_1\ra 1^1S_1 $ & 49.57   & 0.3   & 01.7      & 04.5  & 08.2  & 9.9   & --    & 14.1  & 10.9  & 15.9  & \\
  \hline\hline
  \end{tabular}
  \end{center}
  \end{table*}

\begin{table*}
\caption{Radiative M1 transition widths of $c\bar c$ in (keV)}\label{tab:ccm1}
\begin{center}
{\begin{tabular}{ccccc}
  \hline\hline
  Transition  &  $1^3S_1\to 1^1S_0$ & $2^3S_1\to 2^1S_0$ & $3^3S_1\to 3^1S_0$ & $2^3S_1\to 1^1S_0$\\
  \hline
  ERHM & 0.703 (110) & 0.151 (62)  & 0.023 (17)  & 20.521 (654)\\
  $CPP_{\nu=0.5}$ & 1.86           & 0.03        & 0.004      & 16.52\\
  $CPP_{\nu=1.0}$ & 9.68           & 0.55        & 0.135      & 58.13\\
  $CPP_{\nu=1.5}$ & 20.45          & 2.60        & 0.942      & 108.44\\
  $CPP_{\nu=2.0}$ & 38.35          & 6.92        & 3.241      & 157.23\\
  \cite{nb2006}   & 1.5$\pm$1.0    & --          & --         & --\\
  \cite{tb2005}NR & 2.90 (116)     & 0.21 (48)   & 0.046 (29) &  -- \\
  \cite{ap2010}   & 1.29           & 0.12        & 0.04       & -- \\
   \cite{sfr2007} & 2.7&  1.2      & --          & -- \\
  \cite{pdg2014}  & 1.21$\pm$0.37  & $<$ 0.67    & --         & 3000$\pm$500\\
  \hline\hline
 \end{tabular}}
 \end{center}
\end{table*}

\begin{table*}
\caption{Radiative M1 transition widths of $b\bar b$ in (eV)}\label{tab:bbm1}
\begin{center}
{\begin{tabular}{ccccc}
  \hline\hline
  Transition  & $1^3S_1\to 1^1S_0$ &  $2^3S_1\to 2^1S_0$ & $3^3S_1\to 3^1S_0$ & $2^3S_1\to 1^1S_0$\\
  \hline
  ERHM            & 2.33 (36) & 0.169 (15) & 0.050 (10) & 1395.9 (580)\\
  $CPP_{\nu=0.5}$ & 2.51& 0.01& 0.001& 223.23\\
  $CPP_{\nu=1.0}$ & 9.13& 0.17& 0.036& 799.45\\
  $CPP_{\nu=1.5}$ & 19.12& 0.98& 0.244& 1629.06\\
  $CPP_{\nu=2.0}$ & 31.20& 2.51& 1.088& 2514.04\\
   \cite{ap2010}  & 7.28 & 0.67 &0.19 & -- \\
  \cite{de2003}   & 5.8 (60) & 1.40 (33) & 0.80 (27) & -- \\
  \cite{sfr2007}  & 4.0 &  0.5 & -- & -- \\
  \cite{nby2005}  & 8.95 & 1.51  & 0.826 & --  \\
  \cite{vva2007}  & 9.2 & 0.6 & 0.6 & -- \\
  \cite{tal2003}  & 7.7 (59) & 0.53 (25) & 0.13 (16) & -- \\
  \hline\hline
\end{tabular}}
\end{center}
\end{table*}

\section{Digamma and Digluon Decay Widths}\label{sec:dgg}

Using the model parameters and the radial wave functions, we compute the digamma ($\Gamma_{\gamma\gamma}(\eta_Q)$) and digluon ($\Gamma_{gg}(\chi_Q)$) decay widths. The digamma decay width of the P-wave $Q\bar Q$ state $\chi_{Q1}$ is forbidden according to the Landau-Yang theorem. Most of the quark model predictions for the S-wave $\eta_Q\rightarrow\gamma\gamma$ width are comparable with the experimental result, while the theoretical predictions for the P-wave ($\chi_{Q0,2}\rightarrow\gamma\gamma$) widths differ significantly from the experimental observations \cite{pdg2014}. The contribution from QCD corrections takes care of this discrepancy. The one-loop QCD radiative corrections in the digamma decay widths of $^1S_0 (\eta_Q)$, $^3P_0(\chi_{Q0})$ and $^3P_2(\chi_{Q2})$ are computed using the non relativistic expressions given by  \cite{ap2010,Bhavin2009}:
\be
\Gamma_{\gamma\gamma}(\eta_Q)=\frac{3e_Q^4\alpha_{em}^2 M_{\eta_Q}|R_0(0)|^2}{2m_Q^3} \left[1-\frac{\alpha_s}{\pi}\frac{(20-\pi^2)}{3}\right]\label{gmgme}\ee
\begin{equation}
\Gamma_{\gamma\gamma}(\chi_{Q0})=\frac{27e_Q^4\alpha_{em}^2 M_{\chi_{Q0}}|R_1^{'}(0)|^2}{2m_Q^5}\left[1+B_0\frac{\alpha_s}{\pi}\right]
\label{gmgmc0}\end{equation}
\be
\Gamma_{\gamma\gamma}(\chi_{Q2})=\frac{4}{15}\frac{27e_Q^4\alpha_{em}^2M_{\chi_{Q2}}|R_1^{'}(0)|^2}{2m_Q^5} \left[1+B_2\frac{\alpha_s}{\pi}\right] \label{gmgmc2}\ee
where $B_0=\pi^2/3-28/9$ and $B_2=-16/3$ are the next-to-leading-order (NLO) QCD radiative corrections \cite{Barbieri1981,Kwong1988,Mangano1995}.\\
Similarly, the digluon decay widths of the $\eta_Q$, $\chi_{Q0}$ and $\chi_{Q2}$  states are given by \cite{Lansberg2009}:
\begin{equation}
\Gamma_{gg}(\eta_{Q})=\frac{\alpha_s^2 M_{\eta_Q}|R_0(0)|^2}{3m_Q^3}[1+C_Q(\alpha_s/\pi)]
\label{gamma1}\end{equation}
\begin{equation}
\Gamma_{gg}(\chi_{Q0})=\frac{3\alpha_s^2 M_{\chi_{Q0}}|R_1^{'}(0)|^2}{m_Q^5}[1+C_{0Q}(\alpha_s/\pi)]
\label{gamma2}.\end{equation}
\be
\Gamma_{gg}(\chi_{Q2})=\left(\frac{4}{15}\right)\frac{3\alpha_s^2 M_{\chi_{Q2}}|R_1^{'}(0)|^2}{m_Q^5} [1+C_{2Q}(\alpha_s/\pi)] \label{gamma3}\ee
Here, the quantities in the brackets are the NLO QCD radiative corrections  \cite{Mangano1995} and the coefficients have values of $C_Q=4.4$, $C_{0Q}=10.0$ and $C_{2Q}=-0.1$ for the bottom quark.\\

\section{Radiative E1 and M1 transitions}\label{sec:e1m1}

In the non-relativistic limit, the M1 transition width between two $S$-wave states is given by \cite{nb2006}
\be
\Gamma_{n^3S_1\to n'{^1S_0}\gamma}  = \frac{4}{3}\alpha e_Q^2 \frac{k^3_\gamma}{m^2}  \left | \int\limits_0^\infty r^2dr R_{n'0}(r)R_{n0}(r)j_0(\frac{k_\gamma r}{2})\right|^2, \label{M1NRlimit}\ee
where  $e_Q$ is the fraction of electrical charge of the heavy quark ($e_b  = -1/3$, $e_c=2/3$), $\alpha$ is the fine structure constant and $R_{nl}(r)$ are the radial Schr\"odinger wave functions. The photon energy $k_\gamma$ is nearly equal to the mass difference  of the two quarkonia, so it is of order $mv^2$ or smaller. This is in unlike radiative transitions from a heavy quarkonium to a light meson, such as $J/\psi\to \eta \gamma$, where a hard photon is emitted. Since $r\sim 1/(mv)$, the spherical Bessel function is expanded as
$j_0(k_\gamma r/2) = 1 - (k_\gamma r)^2/24 + \dots$ \cite{nb2006}. While the overlap integral in (\ref{M1NRlimit}) is unity at leading order for $n=n'$ ({\it allowed} transitions), it vanishes for $n\neq n'$ ({\it hindered} transitions). The widths of hindered transitions are determined by higher-order and relativistic corrections only.

In the non-relativistic limit, radiative E1 and M1 transition partial widths are given by \cite{nb2006}

\be
\Gamma_{n^{2S+1}L_{iJ_i} \to n'^{2S+1}L_{fJ_f}\gamma}  =
\displaystyle\frac{4\alpha e_Q^2 k_{\gamma}^3}{3} (2J'+1) max(L_i,L_f)\left \{ \begin{array}{ccc}J_i & 1 & J_f \\ L_f & S & L_i \\ \end{array}\right \} | \langle f | r | i \rangle |^2, \label{E1:NRlimitq}\ee
\be
\Gamma_{n^3S_1 \to n'^1S_0\gamma} =
\frac{4}{3} \frac{2J'+1}{2L+1}\delta_{LL'}\delta_{S,S'\pm 1}\alpha
e_Q^2 \frac{k^3_\gamma}{m^2}\left|\dis\int\limits_0^\infty
r^2dr R_{n'0}(r)R_{n0}(r)j_0(\frac{k_\gamma r}{2})\right|^2 \label{M1:NRlimit}\ee
The CLEO-c experiment has measured the magnetic dipole (M1) transitions $J/\psi(1S) \rightarrow \gamma \eta_c(1S)$ and $\psi(2S) \rightarrow \gamma \eta_c(1S)$ using a combination of inclusive and exclusive techniques reconciling with the theoretical calculations of lattice QCD and effective field theory techniques \cite{rem2009,nb2006}. M1 transition rates are normally weaker than E1 rates, but they are of more interest because they may allow access to spin-singlet states that are very difficult to produce otherwise. The spectroscopic parameters of ERHM and CPP$_{\nu}$ are utilized for the present computations.

\section{Results and conclusions}\label{sec:conclusion}
In this paper, we have employed the masses of the pseudoscalar and vector mesons, their wave functions, and other input parameters from our earlier work \cite{akr2008} for the calculations of the digamma, digluon decay widths as well as E1 $\&$ M1 transitions. E1 and M1 radiative transitions of the $c \bar c$ and $b \bar b$ mesons in the ERHM and Coulomb plus power potential  CPP$_\nu$ models and computed numerical results are tabulated in Tables \ref{tab:ccdigamma}-\ref{tab:bbm1}. The digamma and digluon decay widths of the $c \bar c$ and $b \bar b$ mesons are computed with and without QCD corrections. The ERHM predictions of  digamma decay widths of charmonia for the ground state are found to be comparable to the other theoretical results. In case of the CPP${_\nu}$ model these values are fairly close around $\nu < 1$. A similar trend is found in the case of digluon decay rates of charmonia.  The digamma and digluon decay widths predicted by the  ERHM and CPP${_\nu}$ models are very close to the other theoretical predictions.

The computations of E1 transition widths are done without any relativistic correction terms. This indicates the possible inclusion of the same in the wave function with a single center size parameter. The E1 and M1 transitions of the $c \bar c$ and $b \bar b$ mesons have been calculated by several groups (See Tables \ref{tab:cce1}-\ref{tab:bbm1}) but their predictions are not in mutual agreement. The predictions from References \cite{de2003,sfr2007} and the CPP${_\nu}$ model (at $\nu\simeq 1$ for $c \bar c$ and at $\nu\simeq 1.5$ for $b \bar b$ mesons) are in fair agreement with experimental values. One of the limitations of the CPP$_\nu$ model is the inability to obtain the mass spectra of the  $c \bar c$ and $b \bar b$ mesons at the same potential index $\nu$. The computed magnetic radiative transition rates are tabulated along with other theoretical predictions and available experimental values in Tables \ref{tab:ccm1} and \ref{tab:bbm1}. The values in the parentheses are the energy of the photon in MeV. The transition widths obtained by the potential models show a large deviation from the experimental data; however, the values computed using effective mean field theories ($\Gamma_{J/\psi\to\eta_c\gamma}=1.5\pm 1.0$ keV and $\Gamma_{\Upsilon(1S)\to\eta_b\gamma}=3.6\pm 2.9$ eV \cite{nb2006}), are found to be nearly the same as the potential model results. The photon energies in all the models are found to be nearly the same as the mass splitting. The wide variation in predicted hyperfine splitting leads to considerable uncertainty in the predicted rates for these transitions. Differences in the theoretical assumptions of the potential models make it difficult to draw sharp conclusion about the validity of a particular model because of the lack of experimental data.

{\it J N Pandya acknowledges the financial support extended by University Grants Commission, India for Major Research Project F.No.42-775/2013(SR). A K Rai acknowledges the financial support extended by Department of Science of Technology, India  under SERC fast track scheme SR/FTP/PS-152/2012.}\\

\end{document}